\documentclass[twoside,american,british,english,3p,twocolumn,sort&compress]{elsarticle}
\usepackage[T1]{fontenc}
\usepackage[latin9]{inputenc}
\usepackage{babel}
\usepackage{units}
\usepackage{amsmath}
\usepackage{amssymb}
\usepackage[unicode=true]
 {hyperref}

\makeatletter
 \makeatletter
 \def\ps@pprintTitle{%
   \let\@oddhead\@empty
   \let\@evenhead\@empty
   \let\@oddfoot\@empty
   \let\@evenfoot\@oddfoot
 }
 \makeatother

\makeatother

\begin{document}
\selectlanguage{british}%

\title{Riserratevi sotto coverta ...\foreignlanguage{english}{\tnoteref{t1}}}

\tnotetext[t1]{Shut yourself up below decks ..\foreignlanguage{english}{.}}

\selectlanguage{english}%

\author{Bruno Cocciaro}

\ead{b.cocciaro@comeg.it}

\address{L\foreignlanguage{british}{iceo Scientifico XXV Aprile, Via Milano
2, 56025 Pontedera (Pisa})}
\selectlanguage{british}%
\begin{abstract}
The principle of relativity, as originally expressed by Galileo, points
out that the area of competence of the principle itself is that of
isolated systems as well as inertial reference frames. The principle
does not claim that it is always possible to isolate any physical
system, indeed it leaves it open to the possibility of the existence
of phenomena concerning non isolable physical systems, e.g. phenomena
regulated by some non draggable ether. After the Aspect experiment
realist and local models have been proposed specifically based on
the hypothesis that entangled systems are not isolated. It is hypothesized
that the correlations which allow the violation of Bell\textquoteright{}s
inequality are due to exchanges of superluminal signals between the
various parts of the system and those signals do not generate causal
paradoxes because their propagation is regulated by a non draggable
ether. In the present paper the perfect compatibility of such models
with the relativity theory is strongly advocated. A criterion is finally
proposed to determine the causal ordering between events since, when
there are superluminal signals, that ordering can no longer be associated
to the time ordering induced by the standard synchronization.\end{abstract}
\selectlanguage{english}%
\maketitle

\section{Introduction}

\selectlanguage{british}%
Galileo, \textquotedblleft{}Dialogue concerning the two chief world
system'' \citep{Galileo} (pages 216-217), writes the following passage,
which has become famous as the first statement of the principle of
relativity.
\begin{quotation}
Shut yourself up with some friend in the main cabin below decks on
some large ship, and have with you there some flies, butterflies,
and other small flying animals. Have a large bowl of water with some
fish on it; hang up a bottle that empties drop by drop into a narrow-mouthed
vessel beneath it. With the ship standing still, observe carefully
how the little animals fly with equal speed to all sides of the cabin.
The fish swim indifferently in all direction; the drops fall into
the vessel beneath; {[}...{]} When you have observed all these things
carefully (though there is no doubt that when the ship is standing
still everything must happen in this way), have the ship proceed with
any speed you like, so long as the motion is uniform and not fluctuating
this way and that. You will discover not the least change in all the
effects named, nor could you tell from any of them whether the ship
was moving or standing still. 
\end{quotation}
Over the years the principle of relativity has been presented in different
ways by different scientists. The focus of our work is the importance
to be \textquotedblleft{}below decks\textquotedblright{}. This point
is adequately stressed by Galileo and maybe implied by modern authors
as well; in my opinion, however, it is implicitly overlooked in order
to achieve some results modern scientists almost unanimously agree
on. In the \textquotedblleft{}demonstrations\textquotedblright{} of
the impossibility of superluminal signals (also called tachyons) almost
all modern authors pass over the condition of being \textquotedblleft{}below
decks\textquotedblright{} and these demonstrations are then given,
more or less unconsciously, a validity which goes beyond the hypothesis
on which they themselves are based. The consequence is often a sharp
refusal to take into account any consideration which hypothesizes
the existence of superluminal signals since it is generally believed
that it is proved that those signals give rise to paradoxes, or that
their existence undermines the theory of relativity. Even when confronted
with astonishing experimental results, like those obtained in the
various experiments on the violation of Bell\textquoteright{}s inequality,
e.g. the Aspect test, only a few scientists have agreed with Bell%
\footnote{The book \textquotedblleft{}The Ghost in the Atom\textquotedblright{}
\citep{Bell} is a collection of a series of radio interviews to various
physicists from BBC Radio 3. The passage quoted is taken from one
of the answers given by Bell to the interviewers.%
} \citep{Bell}, stating that ``in these EPR experiments there is
the suggestion that behind the scenes something is going faster that
light\textquotedblright{} (page 49). The prevailing opinion has been,
and still is, that the Aspect test marks the end of the local realism.
This is the amazing conclusion on which there is almost unanimous
agreement since it is believed that the facts themselves (the Aspect
test) compel us to accept it. The main purpose of these pages will
precisely be to demonstrate that it is not compulsory and not even
necessarily preferable from the epistemological point of view, to
interpret the outcome of the Aspect test in the currently dominant
way. If we conveniently emphasize the condition of being \textquotedblleft{}below
decks\textquotedblright{}, which must be expressed in the principle
of relativity, it can be easily shown that the superluminal signals
could exist without generating any paradox or arising any problems
for the theory of relativity; the Aspect test might also find an interpretation
which, depending on personal ideas (or the assumed epistemological
view), could be considered far less amazing than the end of local
realism.

\section{Reference frames and systems of coordinates.}

Before analysing the extract by Galileo quoted above, we must clearly
define what will be meant by \textquotedblleft{}reference frame\textquotedblright{}
in the following pages, specifying the clear distinction with the
concept of \textquotedblleft{}system of coordinates\textquotedblright{}.

We will adopt the definition given by E.Fabri \citep{Fabri} (page
40):
\begin{quotation}
But what is a reference frame? With reference frame I mean an environment,
a laboratory, \emph{a real physical object}, material, really existing,
identified in practical terms. Incidentally, a reference frame must
be rigid: in a lab deforming before my eyes, built like an accordion,
it is difficult to carry out all the measurements or to interpret
them. I will also imply that this laboratory-environment is equipped
with all the measurement instruments I need. {[}...{]} You should
think that in a reference frame, identified as a laboratory, a room,
there are all the instruments you need to take measurements, to carry
out an experiment. This is a reference frame.

Therefore, when I mention two different reference frames you should
think of two such environments separated: that\textquoteright{}s all.
Naturally, nothing prevents me, if, for instance, our reference frame
is this classroom, from agreeing that \emph{x}, \emph{y} and \emph{z}
are respectively the Cartesian coordinates to identify the various
points. However, instead of assuming that SC {[}system of coordinates{]}
I can take any other, without changing the physics for that; only
the numbers expressing the coordinates will change.
\end{quotation}
Briefly, the reference frame, as defined by E. Fabri, is Galileo\textquoteright{}s
\textquotedblleft{}main cabin on some large ship\textquotedblright{}.
And as in Galileo\textquoteright{}s main cabin of the large ship there
are flies, butterflies, a large bowl of water, some fish, a small
bucket, in the reference frame defined by Fabri there are \textquotedblleft{}all
the instruments\textquotedblright{} necessary to carry out any experiment.
Once a reference frame within which a certain experiment is carried
out has been fixed, any system of coordinates can be chosen freely
(for instance, Cartesian, or spherical, cylindrical or others) to
describe the experiment itself. The relativity principle obviously
concerns the reference frames, not the systems of coordinates. Precisely
it concerns the \emph{inertial} reference frames. We should specify
which feature makes a reference frame (a room, a lab) inertial. Galileo
points out also that the relativity principle concerns reference frames
\textquotedblleft{}below decks\textquotedblright{}, we should therefore
also point out which condition should be satisfied in order to state
that a reference frame is \textquotedblleft{}below decks\textquotedblright{}.

\section{\label{sec:Il-principio-di-relativit=0000E0}The relativity principle}

Let\textquoteright{}s now analyse the passage by Galileo in order
to identify the \textquotedblleft{}gist\textquotedblright{} of the
principle he states.\\
Let us focus on what is with a modern expression called \emph{inertiality}
of the reference frame. Galileo defines it assuming that the motion
of the ship \textquotedblleft{}is uniform and not fluctuating this
way and that\textquotedblright{}. In order not to fluctuate this way
and that, the ship must not have interactions with the external world,
or, at least, its \textquotedblleft{}main cabin\textquotedblright{},
which forms our reference frame. We can also say that the sum of all
the interactions the reference frame (the cabin) has with the external
world must be null. For instance, the ship might interact with the
Earth, and, because of that interaction the ship would be subject
to the gravity force, it might then interact with the sea water which
exercises on the ship a force opposite to the gravity force. Naturally,
if we wanted to define as inertial a reference frame having null interaction
with the external world, we might have the problem to define the concept
of \textquotedblleft{}interaction\textquotedblright{}, or, even better,
the concept of \textquotedblleft{}absence of interaction\textquotedblright{};
in my opinion, however, such a concept should be considered as primitive.
I do not deny the importance of discussing about the correct definition
of the concept ``absence of interaction\textquotedblright{}, but
I maintain that this topic goes well beyond the area of competence
we are setting ourselves. Whichever the correct definition of that
concept is, the fact is that the relativity principle concerns inertial
reference frames and that, at least in my opinion, the definition
of inertial reference frame is impossible without using the concept
of \textquotedblleft{}absence of interactions\textquotedblright{}.
Even if we wanted to define as inertial a reference frame where the
first principle of dynamics applies, we would find ourselves in a
situation in which the concept of absence of interaction is considered
as primitive.

Let us now turn to analyse the other point Galileo makes: the inertial
reference frame must be \textquotedblleft{}below decks\textquotedblright{}.
Let\textquoteright{}s hypothesize we are carrying out a certain experiment
in our inertial reference frame. In such an experiment some measurement
instruments as well as some objects which form the system under experimental
study will be involved. I think that the correct interpretation of
the concept \textquotedblleft{}below decks\textquotedblright{} should
be that the external world must not interact with the system under
experimental study. In other words, the system must be \emph{isolated}.
Naturally, during the experiment, our system will be subject to some
interactions; what we hypothesize, however, is that such interactions
find all their causes within our reference frame.

We have therefore an inertial reference frame which, as such, does
not interact with the external world and, consequently, all the measurement
instruments solidal to the reference frame will not receive from the
walls any interaction from the external world. The experiment we are
carrying out inside our inertial reference frame will concern a system
which we hypothesize it is isolated. All interactions occurring during
our experiment will consequently be \emph{internal} to the reference
frame, that is, they will be interactions between parts of the system
or between these and some measurement instrument.

The relativity principle claims that repeating the same experiment
within two different inertial reference frames the same results will
be obtained if the system under our experimental study is \emph{isolated}.

In fact a more extensive form can be given to the relativity principle,
which does not limit it to necessarily isolated systems. Galileo knows
very well that, inside his boat, a privileged direction exists, called
\textquotedblleft{}vertical\textquotedblright{}. It is the direction
of the motion of the bodies which are dropped from a still condition.
When he says that the outcome of two experiments carried out on two
different boats, having a different state of motion with respect to
the sea, will always be the same, he is assuming that the two boats
are placed in the same way with respect to the vertical. We know that
there are various experiments to enable us to understand how the boat
is placed with respect to the vertical. For instance, it is sufficient
to place a small ball on a table, if it does not roll it means that
the surface of the table is perpendicular to the vertical direction.\\
Galileo does \emph{not} claim the physical equivalence of all inertial
reference frames in relation to any one test. Besides, he does not
claim that an \textquotedblleft{}outside\textquotedblright{} with
decisive effects on our experiments (or at least on some of them)
cannot exist. Two boats turned 10 degrees one with the other around
an axis perpendicular to the vertical are surely not equivalent: in
one the ball stays still on the table, in the other it does not.\\
Actually it is not true that Galileo\textquoteright{}s reference frame
is shielded from any possible external influence. Newton revealed
that the privileged direction is given by the interaction of the bodies
with the Earth (i.e. this direction is established by the relation
between our boat and the Earth), however, following what Newton says,
Galileo\textquoteright{}s principle could become:\\
the same experiment must be carried out in two different inertial
reference frames trying to shield them at its best from the effects
that the outside has on the experimental apparatus. Should residual
interactions due to the outside exist, having decisive effects on
the experimental apparatus, all efforts must be made to ensure that
the effects are the same in both reference frames. The principle claims
that only after all these precautions have been taken, will the experiments
carried out in the two reference frames will have the same outcome.\\
Naturally there is the problem of finding a criterion to understand
whether the outside has decisive effects on the experimental apparatus
or not. The relativity principle indicates such a criterion: the same
\emph{operations} are repeated in two different reference frames,
both the operations to prepare the system under experimental study
and the measurement operations; if the measurements have different
results in the two reference frames we will say%
\footnote{The use of this verb shows the epistemological value of the principle
under study (as of all the principles generally employed in physics):
we do not claim to be making a \emph{true} statement, we only state
a modus operandi. Fundamentally we say \textquotedblleft{}Let us try
to interpret the events following this principle''.%
} that the outside has decisive effects on the experimental apparatus,
i.e. it has decisive effects either on the system (at least one of
the two reference frames is not below decks, i.e. it is not isolated)
or on the measurement instruments (at least one of the two reference
frames is \textquotedblleft{}fluctuating this way and that\textquotedblright{}).\\
When we say that the outside has decisive effects on our experimental
apparatus we do not mean non-local interactions. Any experiment always
consists in the preparation of a certain system in a certain way (preparation
of causes), and in the observation of some results of some measurements
made on the system (observation of effects). We say that the system
object of our experimental study is \textquotedbl{}below decks\textquotedbl{}
if the evolution of our system depends exclusively on the causes that
have been prepared by the experimenter. The concept of \textquotedbl{}below
decks\textquotedbl{} does not consist in the mere assertion of the
locality of the interactions (which is implied). The concept defines
systems which are composed only of subsystems that the experimenter
has prepared purposely in the inertial reference frame in which the
experiment takes place. A system is not below decks if its parts may
interact locally, and then internally to the reference frame, with
some subsystem that has not been prepared by the experimenter like,
for example, a field generated by entities that are located \textquotedblleft{}above
decks\textquotedblright{} which has not been properly screened.\\
\emph{This is the origin of the possible non-invariance of the measurements
performed on systems that are not below decks}.\\
But this possible non invariance does not affect in any way the principle
of relativity which states the invariance of the measurements only
for systems that are below decks.\\

The topic (i.e. the identification of the gist of the relativity principle)
could be presented in other terms too. The basic assumption is that
the phenomena, at least those physics deals with, have some causes.
The relativity principle claims that the repetition of the \emph{same}
causes will result in the same effects, regardless of \emph{where}
and \emph{when} the repetition of the causes occurs, that is regardless
of the reference frame inside which the experiment is carried out.\\
For instance, cutting the string of given length which is compressing
a spring of given force constant and given rest length on which the
body \emph{C} of given mass is placed at one end and the body \emph{C'}
of given mass at the other (preparation of the causes), it can be
observed that the measurement of the momentum of the body \emph{C}
will have a certain value \emph{p} in the moment in which \emph{C}
moves away from the spring (effect). It has no relevance which is
the reference frame in which string, spring, \emph{C} and \emph{C'}
have been placed (the reference frame will necessarily have to be
inertial if the measurement instrument is solidal to the reference
frame, as it usually is). What matters is only that the preparation
of the causes is carried out exactly in that way; for instance it
is important that, before cutting the string, there is no relative
motion between string, spring, \emph{C}, \emph{C'} and the instrument
used to measure the momentum of \emph{C} placed where \emph{C} will
move away from the spring%
\footnote{A momentum measurement instrument of body \emph{C} could be a rule,
long \emph{L}, with the direction of the motion of the body (that
is the direction of the spring), equipped with a recorder of instants
$\tau_{in}$ and $\tau_{fin}$, shown on the clock in motion with
\emph{C}, in the moments when the clock (and \emph{C}, too) goes by
the two ends of the rule. We will have $p=m\frac{L}{\tau_{fin}-\tau_{in}}$,
where $m$ is the mass of the body \emph{C}. The correct use of the
instrument requires the instrument to be in rest with respect to the
spring, the string and the body \emph{C} before cutting the string
and not to be subjected to interactions at least until the measurement
procedure has been completed (the only admitted interactions are those
strictly connected to the measurement to take, that is those which
allow the recordings of the instants $\tau_{in}$ and $\tau_{fin}$).%
}. As that effect (the measurement of the value \emph{p}) has only
those causes (string, spring, \emph{C}, \emph{C'} and measurement
instrument prepared in that way), the rest of the universe has no
importance in order to observe that effect. Particularly, the state
of motion of the rest of the world with respect to the system under
study has no importance whatsoever.\\
Probably it will be always impossible to be aware of \emph{all} the
causes of a certain effect. The modus operandi we assume (the relativity
principle) tells us that repeating all the causes we know we should
observe the same effect, it also indicates the possible reasons which
could determine our failure to observe the same effect:

a) a certain cause we are aware of has not been repeated correctly;

b) there is a further cause, inside the reference frame in which the
experiment is carried out, which we do not know;

c) the reference frame is \textquotedblleft{}fluctuating this way
and that\textquotedblright{}, i.e. it is in relation with the outside.
This interaction, passed on to the measurement instruments which,
as prescribed in the instructions for the preparation of the experiment,
are fixed to the reference frame (or in a precise state of motion
relative to the reference frame), makes the outcomes of the measurements
different from those we would have if the reference frame were not
\textquotedblleft{}fluctuating this way and that\textquotedblright{};

d) the reference frame is not \textquotedblleft{}below decks\textquotedblright{},
that means our system has relations with the outside we are not aware
of, consequently, we have neither been able to shield the system from
those interactions nor to prepare it so that such interactions would
repeat always in the same way.\\

The points a) and b) are obvious (so obvious that the relativity principle
does not mention them; however, we could imagine they are implied).
The former simply states that the experiment has been carried out
incorrectly, the latter that the phenomenon under study is not perfectly
known yet%
\footnote{This is essentially always true. The point b) is the one we always
turn to, at least at first, whenever a certain experiment, repeated
after improving the experimental apparatus, produces results different
from those expected.%
}. For instance, in the experiment above, using a sufficiently refined
instrument, we might observe a different \emph{p} from one day to
the next, and then realize that among the possible causes there is
the air humidity rate inside the reference frame.\\
The point c) is well known and states that an experiment prepared
in a non inertial reference frame might have different outcomes from
those obtained from an experiment prepared in an inertial reference
frame.\\
The point d) is the one we are particularly focusing our attention
on. Actually, we are focusing it on the fact that the relativity principle
predicts that \emph{there is also} the point d), that is, it foresees
that there might be different outcomes when repeating the experiment
in two different reference frames even if it were true that:\\
- all the causes we know have been repeated in a perfectly identical
way (point a);\\
- we can show that no cause we are not aware of can exist inside the
two reference frames (point b);\\
- the two reference frames are inertial, as confirmed by the fact
that all the experiments carried out before have always had the same
outcomes in the two reference frames (point c).\\
Galileo claims that, in a situation like the one presented above,
we must not question the validity of the relativity principle yet,
we must not doubt yet that the repetition of the same causes produces
the same effects, on the contrary we must consider the hypothesis
that at least one of the two reference frames may not be \textquotedblleft{}below
decks\textquotedblright{}, that is the outside interacts with our
system and the interaction is different in the two reference frames.\\
Going back to the experiment in our example, we could realize that
rubbing the body \emph{C} two different results are obtained in the
two different inertial reference frames. It might happen that, with
the passing time, in one reference frame the difference between \emph{p}
measured rubbing \emph{C} and \emph{p} measured without rubbing it
will increase, while in another reference frame such a difference
will decrease. Since the two reference frames are inertial (they move
with \textquotedblleft{}free motion\textquotedblright{}, i.e. the
rooms where the experiments are carried out, the two laboratories,
do not interact with the outside; and neither do the measurement instruments
within the labs) and the procedure followed for the preparation of
the experiments in the two laboratories is the same, the relativity
principle require us to go \textquotedblleft{}out there\textquotedblright{}
to look for the causes which might produce different results. Some
confirmations may be found out there. For instance, an electrically
charged body may be found, to which one reference frame is drawing
up while the other is moving away.\\

We conclude this paragraph stressing again our thesis.\\
In many texts it is said that, according to the relativity principle,
it is not possible to identify any experiment which may produce different
outcomes in two different inertial reference frames. For instance
in 1921, in \textquotedblleft{}The meaning of relativity\textquotedblright{}\citep{Einstein(1921)}
Einstein claims (page 14) 
\begin{quotation}
If \emph{K} is an inertial system, then every other inertial system
\emph{K'} which moves uniformly and without rotation relatively to
\emph{K}, is also an inertial system; the laws of nature are in concordance
for all inertial systems. This statement we shall call ``principle
of special relativity.'' 
\end{quotation}
Such a statement is too strong and does not respect the spirit of
the principle originally claimed by Galileo. At least in the opinion
of the writer of these pages. Einstein\textquoteright{}s expression
\textquotedblleft{}moves uniformly and without rotation\textquotedblright{}
is Galileo\textquoteright{}s \textquotedblleft{}the motion is uniform
and not fluctuating this way or that\textquotedblright{}, in Einstein\textquoteright{}s
sentence, however, Galileo\textquoteright{}s \textquotedblleft{}below
decks\textquotedblright{} is missing.\\
We can also note that, in the famous article of 1905, Einstein\citep{Einstein(1905)}
says more cautiously:
\begin{quotation}
Examples of this sort, together with the unsuccessful attempts to
discover any motion of the earth relatively to ``light medium''
suggest that the phenomena of electrodynamics as well as of mechanics
possess no properties corresponding to the idea of absolute rest.
They suggest rather that, as has already been shown to the first order
of small quantities, the same laws of electrodynamics and optics will
be valid for all frames of reference for which the equations of mechanics
hold good. We will raise this conjecture (the purport of which will
hereafter be called the ``Principle of Relativity'') to the status
of postulate {[}...{]} 
\end{quotation}
In the 1905 statement Einstein does not say that \emph{all} natural
laws ``will be valid for all frames of reference for which the equations
of mechanics hold good''. He confines himself to mechanics and electrodynamics,
he will then refer to rules and clocks (saying basically that rules
and clocks behave in the same way in every inertial reference frame).
He will build up the relativity on the basis of these minimal hypothesis.
This does not mean that relativity should be limited to mechanics
and electrodynamics. Every phenomenon which finds its causes within
the reference frame (excluding consequently phenomena ruled by an
\textquotedblleft{}ether\textquotedblright{} in which the reference
frame is immersed) will find a similar description for every reference
frame. However, a phenomenon must not necessarily find all its causes
within the reference frame. A phenomenon must not necessarily be described
in the same way in every reference frame.\\
Definitely, resuming the experiment in our example, the \textquotedblleft{}law
of nature\textquotedblright{} which regulates the motion of the body
\emph{C} rubbed is different in the two reference frames. It becomes
the same only if the two reference frames are shielded from the effects
of external charges, namely if the two reference frames are placed
\textquotedblleft{}below decks\textquotedblright{} also in relation
to experiments carried out with bodies electrically charged.\\
Particularly relevant is the fact that the possible discovery of a
phenomenon which we can not shield from the interactions with the
outside (with the consequence that the \textquotedblleft{}law of nature\textquotedblright{}
for that phenomenon may be different in the different reference frames)
\emph{does not create any problems} for the relativity principle which
indeed predicts that such phenomena may exist and also explains how
to act in such a circumstance (the causes must be looked for \textquotedblleft{}out
there\textquotedblright{}).\\
Even if we did not know how to shield a reference frame from external
electrical fields and were consequently forced to describe the electrical
phenomena differently in each reference frame (as different reference
frames are hit by external electrical fields in theory different one
from the other), still the fact that the physical laws of the electrically
discharged bodies are the same in all inertial reference frames would
apply. The possible finding of a phenomenon for which external causes
to the reference frame were decisive (e.g. a phenomenon with a preferred
reference frame in which the outcomes of the measures associated to
the phenomenon, or at least some of them, are different from the outcomes
of the same measures carried out in other reference frames) \emph{would
not change in any way} the laws of physics which describe the phenomena
with causes within the reference frame.\\
\emph{The possible discovery of phenomena with a preferred reference
frame would not create any problems for the relativity.}\\
In the following pages we will call \emph{inertial} a reference frame
(a room, a lab.) which does not interact with the outside, that is
a reference frame in which the measurement instruments do not have
an interaction with the outside, not even indirectly through their
interaction with the walls of the reference frame which consequently
must not themselves interact with the outside.\\
We will call the following statement the \emph{relativity principle
(PR)}: \\
repeating a certain experiment on a certain system within two inertial
reference frames, the same results will be obtained in both reference
frames if the system under experimental study does not interact with
the outside. The same outcome will be achieved even if the outside
interacted with the system or with the reference frames only if such
interactions are made the same in both reference frames.\\
We will call the following statement the \emph{relativity principle
in its strong form (PRF)}:\\
repeating a certain experiment on a certain system within two inertial
reference frames the same outcome will \emph{always} be obtained in
both reference frames. In order to obtain the same outcome it may
be necessary either to shield the reference frames or to repeat the
possible interactions with the outside of the system under experimental
study in the same way in both the reference frames. It is assumed
that at least one of the two possibilities (the shielding or the identical
repetition in the different reference frames) will always be accessible
by means of operations that could be performed within the reference
frame.\\
The difference between \emph{PR} and \emph{PRF} is clear: the former
admits the possibility that there may be phenomena for which it may
not be possible to be placed \textquotedblleft{}below decks\textquotedblright{}.
The \emph{PR} lets nature decide whether, for a certain phenomenon,
it is possible to be placed below decks (that is it is possible to
work so that the specific phenomenon could have all its causes within
the reference frame, alternatively the possible residual external
causes be repeated in the same way in every reference frame). The
\emph{PRF} states a prescription with which nature should comply:
for any phenomenon it must always be possible either to shield it
from the interactions with the outside which might affect the phenomenon
or to repeat those interactions exactly in the same way in every inertial
reference frame (that means either that the phenomenon, possibly after
the shielding, will find all its causes within the reference frame,
or that the possible external causes from which it can not be shielded
could always be repeated in the same way in every reference frame).

\section{Tachyons and causal paradoxes}

The theorem showing the relation between tachyons and causal paradoxes
is discussed in several texts: e.g. M{\o}ller (1952) \citep{Moeller (1952)}
pages 52-53, Bohm (1996) \citep{Bohm (1996)} pages 186-189, Regge
(1981) \citep{Regge (1981)} page 21, Penrose (1989) \citep{Penrose (1989)}
pages 274-275, Rindler (2006) \citep{Rindler (2006)} page 55.\\
As for all theorems, of course, there is a thesis and there are some
hypothesis on which we base ourselves. The thesis is exactly the paradoxicality
of the tachyons which would allow us to communicate with our past.
For instance, using tachyons, in the afternoon, in the middle of a
storm, we could warn ourselves in the morning to take the umbrella
as it would start to rain around midday. The paradox is that, being
soaked, we could act as to avoid getting wet.\\
We will see that not all the hypothesis of the theorem will be necessarily
proved and that it will be possible to deny the thesis (claiming that
the existence of tachyons is not necessarily paradoxical) leaving
all the physics known so far basically unaltered.\\
In less recent works the theorem is presented incompletely: Einstein
(1907) \citep{Einstein (1907)} pages 379-382, Tolman (1917) \citep{Tolman (1917)}
pages 54-55, Pauli (1921) \citep{Pauli (1921)} page 34, Von Laue
(1922) \citep{Von Laue (1922)} page 70. These \textquotedblleft{}demonstrations\textquotedblright{}
are based on the fact that a tachyon sent from one point to another
in a certain reference frame, if observed from a reference frame $R'$
having adequate motion in relation to \emph{R}, would prove leaving
from a certain point at the instant $t'_{in}$ and arriving at another
definite point at the moment $t'_{fin}<t'_{in}$. This fact is considered
paradoxical, but one forgets that the instants $t'_{in}$ and $t'_{fin}$
(shown on \emph{different} clocks), as all the other instants shown
on any clock in the reference frame under study, depend decisively
on the synchronization procedure conventionally adopted. Consequently
it appears obvious that, on the base of conventional assumptions,
no physically meaningful conclusion can be drawn.\\
The demonstration can be easily completed (as it is done in many recent
works \citep{Bohm (1996),Regge (1981),Penrose (1989),Rindler (2006)})
assuming that we can send from $R'$ to \emph{R} tachyons having the
same characteristics of those sent from \emph{R} to $R'$. This assumption
is considered valid on the base of the relativity principle expressed
in the form called above \emph{PRF}, not in the one originally expressed
by Galileo \citep{Galileo} (\emph{PR}) which, as already said, is
consistent with the principle stated in the form chosen by Einstein
\citep{Einstein(1905)} in 1905.\\
M{\o}ller \citep{Moeller (1952)} represents an exception: he bases
his demonstration on less strong hypothesis than those in \citep{Bohm (1996),Regge (1981),Penrose (1989),Rindler (2006)}.
Since the theorems expressed in \citep{Bohm (1996),Regge (1981),Penrose (1989),Rindler (2006)}
can be obtained as particular case in the demonstration given by M{\o}ller,
we will discuss only this last in detail. It is just thanks to the
extent of the hypothesis assumed that M{\o}ller\textquoteright{}s
demonstration will, at first sight, seem unchallengeable. We will
see that this is not the case.

\subsection{M{\o}ller\textquoteright{}s demonstration}

M{\o}ller considers two reference frames, \emph{S} and $S'$, in
uniform relative motion, where coordinate axis have been defined $\left(x,y,z\right)$
and $\left(x',y',z'\right)$ in the usual way, and where the clocks
have been synchronized standardly. The motion of $S'$ with respect
to \emph{S} is defined by the speed $\left(v,0,0\right)$. Both the
clocks fixed at the origin of the two reference frames show the instant
0 when the two origins are overlapped. The link between the coordinates
defined in the two reference frames is then given by Lorenz\textquoteright{}s
transformations in their usual form:
\[
\left\{ \begin{array}{l}
x=\gamma\left(x'+vt'\right)\\
t=\gamma\left(t'+vx'/c^{2}\right)\\
y=y'\\
z=z'
\end{array}\right.
\]
where $\gamma=\left(1-\nicefrac{v^{2}}{c^{2}}\right)^{-\frac{1}{2}}$
with \emph{-c<v<c}, being \emph{v} the speed of $S'$ with respect
to \emph{S}.\\
M{\o}ller hypothesizes that, in the moment when the origins of the
two reference frames overlap, a tachyon with speed $u'$\emph{>c}
in $S'$ can be sent out, from the origin, in the negative direction
of the axis $x'$. It should be noted that, no hypothesis is made
about the motion of the emitter of the tachyon. We do not hypothesize
that the emitter is fixed in $S'$ or in \emph{S}. We only hypothesize
that, somehow, this tachyon has been sent out from that point towards
that direction. And we simply say that, if the signal sent out is
a tachyon, it will be seen like that in $S'$. The fact that, if a
signal proves superluminal in a reference frame, then it will prove
superluminal in every reference frame, comes from the invariance of
the difference $dt^{2}-dx^{2}/c^{2}$: if in one reference frame $dt^{2}-dx^{2}/c^{2}<0\Leftrightarrow\left(\nicefrac{dx}{dt}\right)^{2}>c^{2}$,
then in every reference frame it will be $u'^{2}=\left(\nicefrac{dx'}{dt'}\right)^{2}>c^{2}$.
This last equation is considered equivalent to $u'$>\emph{c} since
the sign of $u'$ is thought to be given by the direction of propagation
of the tachyon. Basically we say that, if, when the clock fixed at
the origin shows the instant 0, it is possible to send out a tachyon
from the origin in the negative direction of the $x'$ (and we hypothesize
it is possible) then the equation describing the motion of that tachyon,
in the reference frame $S'$, will be $x'=-u't'$ with $u'$\emph{>c}
(positive).\\
At this point M{\o}ller says that the tachyon will arrive in the
point \emph{P}, associated to the abscissa $x'_{p}<0$ in the $x'$
axis, in the moment when the clock fixed in $x'_{p}$ shows the instant
$t'_{1}>0$, with $x'_{p}=-u't'_{1}$. So Lorentz\textquoteright{}s
transformations give the space time coordinates associated with the
event \textquotedblleft{}arrival of the tachyon in \emph{P}\textquotedblright{}
in the reference frame \emph{S} (eqs. (50) in \citep{Moeller (1952)}):
\begin{equation}
\left\{ \begin{array}{l}
x_{p}=\gamma\left(x'_{p}+vt'_{1}\right)=-\gamma t'_{1}\left(u'-v\right)<0\\
t_{1}=\gamma\left(t'_{1}+vx'_{p}/c^{2}\right)=\gamma t'_{1}\left(1-vu'/c^{2}\right)
\end{array}\right..\label{eq:1}
\end{equation}
We now hypothesize that, soon after the arrival of the tachyon in
\emph{P}, a new tachyon is sent out from \emph{P} towards \emph{O}
and that the speed of this latter tachyon is \emph{w>c} with respect
to \emph{S}. As before, we could say that for this second tachyon
too only the bare essential has been assumed: the reference frame
\emph{S} must see this second signal as superluminal. Thus, as the
first hypothesis seemed summarizable as \textquotedblleft{}tachyons
can be sent out in the negative direction of the axis $x'$\textquotedblright{}
this seems summarizable as \textquotedblleft{}tachyons can be sent
out in the positive direction of the axis \emph{x}\textquotedblright{}.\\
If we laid down \emph{w}=$u'$ justifying the position on the account
of the \emph{PRF} (if there are tachyons which $S'$ sees moving at
the speed $u'$ towards \emph{S}, then there must be tachyons which
\emph{S} sees moving at the speed \emph{w=}$u'$ towards $S'$) then
the demonstration would become equivalent to the one given in \citep{Bohm (1996),Regge (1981),Penrose (1989),Rindler (2006)},
but M{\o}ller does not fix \emph{w=}$u'$, thus making his demonstration
more general. M{\o}ller\textquoteright{}s hypothesis would simply
appear that, given any direction, it is possible to send out tachyons
along that direction, in both ways.\\
The demonstration goes on by clarifying the equation of motion of
the second tachyon in the variables \emph{x}, \emph{t}, defined in
\emph{S} (eq. (51) in \citep{Moeller (1952)}):
\[
x=w\left(t-t_{1}\right)+x_{p}.
\]
From this equation, employing \eqref{eq:1}, we will obtain that the
second tachyon will reach \emph{O} (\emph{x}=0) when the clock fixed
in \emph{O} shows the instant $t_{2}$ given by (eq. (52) in \citep{Moeller (1952)}):
\[
t_{2}=t_{1}-x_{p}/w=\gamma t'_{1}\left(1-\frac{u'v}{c^{2}}+\frac{u'-v}{w}\right).
\]
We then have that the first tachyon leaves \emph{O} when the clock
fixed in \emph{O} shows the instant 0 and the second tachyon (whose
departure from \emph{P}, we could imagine, has itself been caused
from the detection in \emph{P} of the first tachyon) reaches \emph{O}
when the clock fixed in \emph{O} shows the instant $t_{2}$. If it
could be $t_{2}<0$ we would have a clear causal paradox.\\
Altering slightly the last step of the demonstration given by M{\o}ller%
\footnote{M{\o}ller reaches the thesis $t_{2}<0$, hypothesizing that $u'>\nicefrac{c^{2}}{v}$
and $w>\nicefrac{\left(u'-v\right)}{\left(\nicefrac{u'v}{c^{2}}-1\right)}$
are chosen (eq. (53) in \citep{Moeller (1952)}), but, in my opinion,
doing so it is not clear if the thesis is to be understood as always
valid, i.e. regardless of the velocities of the tachyons hypothesized,
subject to the adequate choice of \emph{v}.%
}, we notice that 
\begin{equation}
v>\frac{u'+w}{1+\frac{u'w}{c^{2}}}\Leftrightarrow\left(1-\frac{u'v}{c^{2}}+\frac{u'-v}{w}\right)<0\label{eq:2}
\end{equation}
and being $t'_{1}>0$ we have that if \emph{v}, the speed of $S'$
with respect to \emph{S}, were higher than $\nicefrac{\left(u'+w\right)}{\left(1+\nicefrac{u'w}{c^{2}}\right)}$,
we would then have the paradoxical thesis $t_{2}=\gamma t'_{1}\left(1-\nicefrac{u'v}{c^{2}}+\nicefrac{\left(u'-v\right)}{w}\right)<0$.
As for every $u'>c$ and \emph{w>c} we always have $\nicefrac{\left(u'+w\right)}{\left(1+\nicefrac{u'w}{c^{2}}\right)}<c$,
the paradoxical thesis would appear inevitable, subject to adequate
choice of \emph{v}.

\subsubsection{Analysis of M{\o}ller\textquoteright{}s demonstration}

Between the lines of M{\o}ller\textquoteright{}s demonstration there
is a hypothesis which could be missed out if not stressed properly.
Saying that it will always be possible to choose \emph{v} higher than
$\nicefrac{\left(u'+w\right)}{\left(1+\nicefrac{u'w}{c^{2}}\right)}$
the fact that the rise of \emph{v} may be uninfluential on the values
of $u'$ and \emph{w} is taken for granted. Actually, we take for
granted that, if $u'$ and \emph{w} depended on \emph{v}, this dependence
would be such to make it anyway possible to choose an adequate $v\in\left(-c,c\right)$
so to satisfy the \eqref{eq:2}. This hypothesis has physical significance,
it describes the behaviour of nature. Naturally, it is correct to
assume hypothesis of physical significance and to find their consequences,
but it is important to be well aware of that. If we wanted to deny
the consequences, we would know which physical hypothesis we should
reject. In our specific case, if we wanted to deny the thesis of the
paradoxicality of the superluminal signals, we should prove the existence
of alternative physical hypothesis on the basis of which the \eqref{eq:2}
proves always unsatisfied.\\
A possible alternative hypothesis is the existence of a preferred
reference frame, $S_{0}$, which supports the propagation of the tachyons,
as the air supports the propagation of sound signals. Once the clocks
in $S_{0}$ have been synchronized according to the standard relation,
the velocity of a tachyon in the reference frame $S_{0}$ will be
always the same, regardless of the direction of propagation. We will
call that velocity $\beta_{t}c$, with $\beta_{t}>1$. The existence
of several types of tachyons, associated to different values of $\beta_{t}$
could be supposed; however, all of them will have the same reference
frame $S_{0}$ as support. Once called $\beta c$ the velocity of
\emph{S} with respect to $S_{0}$, the velocity of a tachyon with
respect to a general reference frame \emph{S} could be obtained from
the known law of the speed composition (which is a direct consequence
of Lorentz\textquoteright{}s transformations).

Finally, let us now analyse M{\o}ller\textquoteright{}s demonstration
considering our alternative hypothesis. We call $\beta c$ e $\beta'c$
respectively the velocities of \emph{S} and $S'$ in relation to $S_{0}$.
We also say that the first tachyon (the one with the equation of motion
$x'=-u't'$ in $S'$) has the speed $\beta_{t}c$ in $S_{0}$ (that
is its equation of motion in $S_{0}$ will be $x=-\beta_{t}ct$),
whereas the second tachyon (the one with equation of motion $x=w\left(t-t_{1}\right)+x_{p}$
in \emph{S}) has the velocity $\bar{\beta}_{t}c$ in $S_{0}$. From
the law of the speed composition quoted above we obtain: 
\begin{equation}
\frac{u'}{c}=\frac{\beta_{t}+\beta'}{1+\beta'\beta_{t}}\label{eq:3}
\end{equation}
 and 
\begin{equation}
\frac{w}{c}=\frac{-\bar{\beta}_{t}+\beta}{\beta'\bar{\beta}_{t}-1}.\label{eq:4}
\end{equation}
 The speed of $S'$ with respect to \emph{S} will be equal to 
\begin{equation}
v=\frac{\beta'-\beta}{1-\beta'\beta}c.\label{eq:5}
\end{equation}
Using \eqref{eq:3}-\eqref{eq:5}, \eqref{eq:2} can be written in
the form $\frac{v}{c}>\frac{\frac{v}{c}\left(1+\beta_{t}\bar{\beta}_{t}\right)+\left(\beta_{t}+\bar{\beta}_{t}\right)}{\frac{v}{c}\left(\beta_{t}+\bar{\beta}_{t}\right)+\left(1+\beta_{t}\bar{\beta}_{t}\right)}$
equivalent to
\begin{equation}
\frac{\left(\beta_{t}+\bar{\beta}_{t}\right)\left[1-\left(\frac{v}{c}\right)^{2}\right]}{\frac{v}{c}\left(\beta_{t}+\bar{\beta}_{t}\right)+\left(1+\beta_{t}\bar{\beta}_{t}\right)}<0.\label{eq:6}
\end{equation}
The numerator of the left member in the inequality \eqref{eq:6} proves
trivially positive for every $v\in\left(-c,c\right)$. The sign of
the denominator is easily obtained noting that $\left(\nicefrac{v}{c}\right)\left(\beta_{t}+\bar{\beta}_{t}\right)+\left(1+\beta_{t}\bar{\beta}_{t}\right)>0\Leftrightarrow\nicefrac{\left(1+\beta_{t}\bar{\beta}_{t}\right)}{\left(\beta_{t}+\bar{\beta}_{t}\right)}>-\left(\nicefrac{v}{c}\right)$,
inequality which is satisfied for every $v\in\left(-c,c\right)$,
being $\nicefrac{\left(1+\beta_{t}\bar{\beta}_{t}\right)}{\left(\beta_{t}+\bar{\beta}_{t}\right)}>1$
if, as in our case, $\beta_{t}>1$ and $\bar{\beta}_{t}>1$. Since
both numerator and denominator are positive, we obtain that \eqref{eq:6}
is never satisfied, it means that, assumed our alternative hypothesis,
it is not possible to determine any $v\in\left(-c,c\right)$ which
makes the situation presented by M{\o}ller paradoxical. The tachyons
ruled by a preferred reference frame which supports their propagation
do not arise any causal paradox.

\section{Local realism}

Let us analyse some other extracts from the interview to Bell already
mentioned \citep{Bell} (pages 48-50):
\begin{quotation}
Question (P. C. W. Davies \& J. R. Brown): \emph{Bell's inequality
is, as I understand it, rooted in two assumptions: the first is what
we might call objective reality - the reality of the external world,
independent of our observations; the second is locality, or non-separability,
or no faster-than-light signalling. Now, Aspect's experiment appears
to indicate that one of these two has to go. Which of the two would
you like to hung on to?}

Answer (J. Bell): Well, you see, I don't really know. For me it's
not something where I have a solution to sell! For me it's a dilemma.
I think it's a deep dilemma, and the resolution of it will not be
trivial; it will require a substantial change in the way we look at
things. But I would say that the cheapest resolution is something
like going back to relativity as it was before Einstein, when people
like Lorentz and Poincaré thought that there was an aether - a preferred
reference frame - but that our measuring instruments were distorted
by motion in such a way that we could not detect motion through the
aether. Now, in that way you can imagine that there is a preferred
frame of reference, and in this preferred frame of reference things
do go faster than light. But then in other frames of reference they
seem to go not only faster that light but backwards in time, that
is an optical illusion.

Q: \emph{Well, that seems a very revolutionary approach!}

A: Revolutionary or reactionary, make your choice. But that is certainly
the cheapest solution. Behind the apparent Lorentz invariance of the
phenomena, there is a deeper level which is not invariant.

Q: \emph{Of course the theory of relativity has a tremendous amount
of experimental support, and it's hard to imagine that we actually
go back to a pre-Einstein position without contradicting some of this
experimental support. Do you think it's actually possible?}

A: Well, what is not sufficiently emphasized in textbooks, in my opinion,
is that pre-Einstein position of Lorentz and Poincaré, Larmor and
Fitzgerald was perfectly coherent, and is not inconsistent with relativity
theory. The idea that there is an aether, and these Fitzgerald contractions
and Larmor dilations occur, and that as a result the instruments do
not detect motion through the aether - that is a perfectly coherent
point of view.

Q: \emph{And it was abandoned on grounds of elegance?}

A: Well, on the ground of philosophy; that what is unobservable does
not exist. And also on grounds of simplicity, because Einstein found
that the theory was both more elegant and simpler when we left out
the idea of the aether. {[}...{]} The reason I want to go back to
the idea of an aether here is because in these EPR experiments there
is the suggestion that behind the scenes something is going faster
than light. Now, if all Lorentz frames are equivalent, that also means
that things can go backward in time.

Q: \emph{Yes, and that is the big problem.}

A: It introduces great problems, paradoxes of causality and so on.
And so it's precisely to avoid these that I want to say there is a
real causal sequence which is defined in the aether. Now, the mystery
is, as with Lorentz and Poincaré, that this aether does not show up
at the observational level. It is as if there is some kind of conspiracy,
that something is going on behind the scenes which is not allowed
to appear on the scenes. And I agree that that's extremely uncomfortable.

Q: \emph{I'm sure Einstein would turn in his grave!}

A: Absolutely. And that's very ironic, it is precisely his own theory
of relativity which creates difficulties for this interpretation of
quantum theory (which is in the spirit of Einstein's unconventional
view of quantum mechanics). 
\end{quotation}
I would say that, from the extract quoted above, we can infer that
both Bell and Davies agree on the fact that the introduction of a
preferred reference frame would result in a return to the pre-relativity
period. This would, in their opinion, make Einstein turn in his grave!
The essence of what said above in paragraph \ref{sec:Il-principio-di-relativit=0000E0}
is that the thesis of the incompatibility between relativity and the
existence of a preferred reference frame proves groundless. Although
this thesis has an almost unanimous support in the scientific community,
in my opinion, it should be completed with arguments aiming at demonstrating
that relativity should be based on \emph{PRF} rather than on the more
general \emph{PR}. I personally do not see those arguments.

There is another reason why Bell defines \textquotedblleft{}extremely
uncomfortable\textquotedblright{} his \textquotedblleft{}cheapest
resolution\textquotedblright{}: Bell maintains that the aether he
proposes would be non observable, like the one hypothesized by Lorentz
and Poincaré for the electromagnetic phenomena. We will now turn to
discuss this alleged non observability.

\subsection{\label{sub:Eberhard}Eberhard}

In 1989 P. H. Eberhard presented a work \citep{Eberhard89} (see also
\citep{Eberhard93}) in which he develops Bell\textquoteright{}s ideas
expressed above giving them a much more concrete substance than what
might result from a radio interview.\\
What, according to Bell, is \textquotedblleft{}something (...) going
faster than light\textquotedblright{} becomes for Eberhard a ``collapse
operator'' which is always generated locally where (and when) a measurement
is taken. The measurement is to be taken on a particle belonging to
a multi-particle system in an entangled state. Then the collapse operator
propagates at superluminal speed and the correlations of the quantum
mechanics only exist if the measurements on the other particles are
made after these have been reached by the collapse operator. Essentially,
the collapse operator is the tachyon, or the tachyonic wave, which
\textquotedblleft{}informs\textquotedblright{} the other entangled
particles (or the measurements devices involved with the other entangled
particles) about the outcome of the first measurement. The other particles
(or the corresponding measurements devices), following the interaction
with the tachyon, will have such a state to make the measurements
appear correlate according to the predictions of the quantum mechanics.
For sufficiently fast measurements the tachyon might not succeed in
\textquotedblleft{}informing\textquotedblright{} all the other particles
in due time, consequently, there would be a decrease in the correlations
foreseen by the quantum mechanics. Since the measurement events of
different particles are, or can generally be, space-like, assumed
the standard synchronization in every reference frame, the time order
of the measurement events will not be the same in all reference frames:
in some of them, some tachyons will travel \textquotedblleft{}backwards
in time\textquotedblright{}. Because of that, Eberhard talks about
Lorentz non-invariance of his model. The \textquotedblleft{}correct\textquotedblright{}
order of the events is established by the time sequence in the reference
frame (which will become preferred for this reason) in which the collapse
operator propagates isotropically, that is the reference frame where,
once the clocks have been synchronized according to the standard relation,
the velocity of the tachyons is the same in every direction. In other
reference frames the time sequence could be reversed. What Bell calls
\textquotedblleft{}optical illusion\textquotedblright{} becomes in
Eberhard the non-invariance for Lorentz\textquoteright{}s transformations
of his own model. We will get back below to the discussion of this
so-called \textquotedblleft{}non-invariance\textquotedblright{}. Now
we will focus our attention on a crucial aspect of the model, i.e.
on the \emph{observability} of the preferred reference frame, quoting
directly from the conclusions of Eberhard\textquoteright{}s work mentioned
above (\citep{Eberhard89} pages 204-205):
\begin{quotation}
In the model, there is a privileged space-time rest frame and a parameter
\emph{V} with the dimension of a velocity; \emph{V} is assumed to
be much larger than \emph{c}. If a model of this type is correct,
no violation of the prediction of quantum theory is expected if time
intervals between measurements are large, i.e., larger than the time
necessary to travel between the locations of these measurements at
the velocity $V>c$ in the privileged rest frame. In particular, no
identification of the privileged rest frame would be possible using
such measurements. However, by performing multiple-measurement experiments
with large distances $\Delta x$ and time intervals smaller than $\nicefrac{\Delta x}{V}$
between measurements, violation of quantum theory may be expected.
Some identification of the privileged rest frame may become possible.

If value of \emph{V} is very high, the time interval $\nicefrac{\Delta x}{V}$
is so small that the multiple-measurement experiments mentioned above
become impractical. As long as this is true, the model will show no
observable discrepancy with quantum theory. Its impact will be primarily
philosophical.

If parameter \emph{V} is not too large, multiple-measurement experiments
of the type described above may reveal a violation of quantum theory
and of Lorentz invariance at the same time. Whenever such experiments
uphold the predictions of quantum theory, they allow one to derive
a lower limit for the parameter \emph{V}. 
\end{quotation}
Called $\vec{v}$ the speed of the preferred reference frame with
respect to the Earth and $v_{t}$ the parameter called \emph{V} by
Eberhard above, that is the module of the speed of the tachyons in
the preferred reference frame, using Lorentz\textquoteright{}s transformations,
the events in the reference frame of the laboratory can be described
and, for the given $\vec{v}$ and $v_{t}$, it will be possible to
evaluate the features the experimental apparatus will have to have
in order to highlight the loss of the correlation predicted by the
model \citep{Salart,Cocciaro}. The details of the references \citep{Salart,Cocciaro}
show how we can trace back the measurements of $\vec{v}$ and $v_{t}$
from the observation of these losses of correlation.

As stressed above by Eberhard, whatever the features of an experimental
apparatus are, the model could not be considered falsified in case
no loss of correlation should be observed (only a minimum value for
$v_{t}$ could be fixed, which, however, will prove dependent on $\left|\vec{v}\right|$).
This could make somebody consider the model not scientific as not
falsifiable. Personally, I cannot agree with this opinion as it would
be as considering not scientific the hypothesis of spherical Earth,
or of light propagating with finite speed, in view of the experimental
evidence which, within the limits of measurement uncertainties, is
consistent with the hypothesis of flat Earth or of light propagating
at infinite speed respectively. Mention could be made that similar
remarks were moved to the Copernican theory: if the Earth moved, phenomena
of parallax with the stars should be observable. Galileo's answer%
\footnote{``I wish have said that if such a variation were perceived, nothing
would remain that could cast doubt upon the earth's mobility, since
no counter could be found to such an event. But even thought this
may not make itself visible to us, the earth's mobility is not thereby
excluded, nor its immobility necessarily proved. It is possible, Copernicus
declares, that the immense distance of the starry sphere makes such
small phenomena unobservable. And as has already been remarked, it
may be that up to the present they have not even been looked for,
or, if looked for, not sought out in such a way as they need to be;
that is with all necessary precision and minute accuracy.'' \citep{Galileo}
(pag. 489)%
}, quoting Copernicus, was that because of the big distance of stars
from the Earth, such phenomena were unobservable with the instruments
available at that time; however, in principle they could have be observed
with measuring instruments endowed with better resolution, as it actually
happened later: in 1839 F. W. Bessel \citep{Bessel} published the
first measurement on the annual parallax of a star (Cygni 61). At
the time Galileo's answer probably sounded weak, considering also
that neither Copernicus nor Galileo were able to estimate the resolution
necessary to observe those phenomena of parallax, but it was the only
possible one. The situation is almost identical for the model under
study: only after $\vec{v}$ and $v_{t}$ are know will it be possible
to estimate the minimum sensitivity an experimental apparatus must
have in order to be able to falsify the quantum mechanics. Before
then, it will be possible only to carry out increasingly more sensitive
experiments which will only raise the minimum value of $v_{t}$ consistent
with the experiments which will remain consistent with the quantum
mechanics.\\
Anyway, independently from the opinions we may have on what should
be considered \textquotedblleft{}scientific\textquotedblright{}, it
could be useful to keep in mind that the model has already attracted
the attention of some experimental groups as well as prestigious magazines.
A first test \citep{Scarani}, carried out in 2000, limited the experimental
analysis only to some directions of the vector $\vec{v}$, problem
overcome in later works \citep{Salart,Cocciaro}. None of the three
experiments has revealed any loss of correlation. The minimum value
of $v_{t}$ that can be deduced from those experiments ranges between
$0.6\cdot10^{4}\, c$ and $1.8\cdot10^{5}\, c$ for $\left|\vec{v}\right|<0.1\, c$,
depending on the value of $\left|\vec{v}\right|$. For $\left|\vec{v}\right|>0.1\, c$
the limit value is inferior to $0.6\cdot10^{4}\, c$ and tends to
$c^{+}$ for $\left|\vec{v}\right|\rightarrow c^{-}$, i.e., fixing
the sensitivity of an experimental apparatus, the losses of correlation
would be unobservable even if it were $v_{t}\gtrsim c$ if $\left|\vec{v}\right|$
were sufficiently close to \emph{c} \citep{Cocciaro}.

\subsection{Bohm and Hiley}

In 1993 D. Bohm and B. J. Hiley \citep{Bohm} published a book in
which there is a model which follows the main aspects (and faces the
same problems) of the model presented some years before by Eberhard.
The following extract is taken from pages 292-294:
\begin{quotation}
For example, one could suppose that in addition to the known types
of field there was a new kind of field which would determine a space-like
surface along which nonlocal effects would be propagated instantaneously.
At present we can say very little about this field, but one could
surmise that this space-like surface would be close to a hyperplane
of constant time as determined in a certain Lorentz frame. A good
candidate for such a frame could be obtained by considering at each
point in space-time, the line connecting it to the presumed origin
of the universe. This would determine a unique time order for the
neighbourhood of that point around which one would expect isotropic
properties in space. We may plausibly conjecture that this frame would
be the one in which the 3K background radiation in space has an isotropic
distribution.

This unique frame would not only make possible a coherent account
of nonlocal connections, but could also be significant in other ways.

{[}...{]} This brings us to the second objection, i.e. that the absolute
frame is unobservable. Certainly if we restrict ourselves to the assumptions
that have been made thus far, this would be a valid criticism. However,
in terms of our ontological approach it is possible to alter these
assumptions in such a way that the absolute frame would be observable,
while negligible changes would be produced in the domain of experiments
that have been possible thus far.

{[}...{]} Let us assume then that the long range connections of distant
systems are not truly nonlocal, as is implied by the quantum theory,
but that they are actually carried in the preferred frame at a speed
that is finite, but very much greater than that of light. For measurements
made at levels of accuracy thus far available, the results will be
very close to those predicted by the present quantum theory. But if
we can make measurements in periods shorter than those required for
the transmission of quantum connections between particles, the correlations
predicted by quantum theory will vanish. In effect we would thus be
explaining quantum nonlocality as an outcome of a deeper kind of non-Lorentz
invariant locality. This however is relevant only in a domain beyond
that in which current quantum theory is adequate.

If we consider such modifications of the theory, it then becomes possible
to contemplate an experimental test that would reveal the preferred
frame. In essence this test would consist of doing an EPR experiment
in which the relative time of detection of the two particles was extremely
accurately determined. In principle the Bell inequality would then
no longer be violated for there would be no time for the disturbance
of one particle to propagate to the other before the measurement was
made on it. And from this it follows that the latter particle would
no longer go into a corresponding state of close correlation with
the first one. Of course this will require a measurement of extremely
high accuracy because the speed of transmission of the quantum connection
is assumed to be very much greater than that of light. It is clear
that the accuracy required is far beyond that of the Aspect experiment.

In certain ways this test is reminiscent of the Michelson-Morley experiment,
but it differs in the crucial respect that what is at issue here is
not the measurement of the speed of light, but rather a measurement
of the immensely greater speed of propagation of our assumed quantum
connection of distant particles. Nevertheless as in the Michelson-Morley
experiment, we would have to take into account the speed of the earth
relative to the preferred frame. Thus one would have to make measurements
in different directions and at different times. If one detected a
change between these measurements, one would be able to demonstrate
the existence of a preferred frame thus making the latter in principle
observable.

However, if such changes were found, this would indicate a failure
of both quantum mechanics and relativity, which would be much more
significant than the mere observation of the speed of the preferred
frame. The meaning of such a result would be that we can eliminate
quantum nonlocality provided that we assume a deeper level of reality
in which the basic laws are neither those of quantum theory nor of
relativity (which latter come out as suitable limiting cases and approximations). 
\end{quotation}
The measurements \textquotedblleft{}of extremely high accuracy\textquotedblright{}
Bohm and Hiley speak about are clearly similar to those presented
in the experiments \citep{Salart,Cocciaro,Scarani}, which we have
already presented in the paragraph \ref{sub:Eberhard}. We can see
that Bohm and Hiley too point out that, in principle, such experiments
make the preferred reference frame they propose observable.

The problem which brings together Eberhard, Bohm and Hiley is the
so-called Lorentz non-invariance of the model. Undoubtedly, if we
hypothesize phenomena physically \emph{different} in a certain reference
frame in respect with all the others, we can\textquoteright{}t expect
to have the same descriptions of those phenomena for each reference
frame. But what is at stake here is not a descriptive issue. It is
the substantial issue mentioned above, which we will now at last deal
with.

\subsection{Lorentz invariance}

How the model is non invariant for Lorentz\textquoteright{}s transformations
is clearly explained by the authors mentioned above. For instance,
Eberhard \citep{Eberhard89} says (pages 178-179): 
\begin{quotation}
The sequence of collapses matches the time order of the measurements
in the chosen space-time rest frame. When measurements are performed
at different points in space outside the light cone of one another,
that time order depends on the rest frame chosen. In this sense, these
formalisms are not Lorentz invariant. 
\end{quotation}
As I already said above, the real issue does not concern formalisms.
Actually, it concerns a formalism with a much heavier relevance than
this \textquotedblleft{}Lorentz invariance\textquotedblright{}. The
question is: does the model challenge the theory of relativity?

We have already seen that the question is asked by Davies and Brown
as soon as Bell hypothesizes a ``deeper level which is not invariant'':
\begin{quotation}
``Of course the theory of relativity has a tremendous amount of experimental
support, and it's hard to imagine that we actually go back to a pre-Einstein
position without contradicting some of this experimental support.
Do you think it's actually possible''?
\end{quotation}
However, in my opinion, Bell\textquoteright{}s answer is definitely
weak. Basically Bell advocates the return to \textquotedblleft{}pre-Einstein\textquotedblright{}
positions. Bell does not answer that \textquotedblleft{}the tremendous
amount of experimental support\textquotedblright{} consists entirely
of experiments carried out \textquotedblleft{}below decks\textquotedblright{},
i.e. on isolated systems, whereas the ``deeper level which is not
invariant\textquotedblright{} aims at reminding us that it is not
sure that a certain phenomenon has necessarily all its causes below
decks. It is not necessarily true that any system is isolated. A system
of particles in an entangled state might interact with a non draggable
ether from the reference frame in which the system is under experimental
study %
\footnote{Were it possible to drag the ether which is assumed to be the support
of the propagation of the tachyons, then the demonstrations on the
causal paradoxes would prove valid. My thanks to V. Moretti for this
important remark on the non draggability of the ether hypothesized
by the above model, which he expressed in a public discussion on usenet.%
}.

Bohm and Hiley have no doubts about the issue, either; they say clearly
that, in their view, the model challenges the theory of relativity:
\begin{quotation}
The most essential feature of special relativity is Lorentz invariance
(\citep{Bohm} page 289);

However, if such changes were found {[}that is if the losses of correlation
predicted by the model were observed{]}, this would indicate a failure
of both quantum mechanics and relativity, which would be much more
significant than the mere observation of the speed of the preferred
frame. The meaning of such a result would be that we can eliminate
quantum nonlocality provided that we assume a deeper level of reality
in which the basic laws are neither those of quantum theory nor of
relativity (which latter come out as suitable limiting cases and approximations)
(\citep{Bohm} page 294).
\end{quotation}
Eberhard sounds slightly more cautious. He does not explicitly mention
the violation of the theory of relativity (\citep{Eberhard89} page
170):
\begin{quotation}
The concept of locality used here is different from the Lorentz-invariant
concepts of locality that led to the EPR paradox, because these latter
concepts are incompatible with superluminous effects. I believe this
rudimentary concept of locality contains all the locality requirements
that could have been expected from a theory before relativity was
discovered. The kind of locality aimed at here is the same kind of
locality obtained at the time the Maxwell equation were written. 
\end{quotation}
Indeed, that \textquotedblleft{}before relativity was discovered\textquotedblright{}
seems to imply that, in Eberhard\textquoteright{}s opinion, his \textquotedblleft{}rudimentary
concept of locality\textquotedblright{} is in contrast with relativity
and that it should be in contrast because it results in a non Lorentz
invariant model.

Surely Eberhard does not support the argument strongly advocated in
these pages, namely that \emph{relativity does not require all phenomena
to be necessarily referable to Lorentz invariant descriptions}.

We could wonder what would be left of relativity if we took away what,
according to Bohm and Hiley, is its \textquotedblleft{}most essential
feature\textquotedblright{}. I think it would be left what relativity
has always been. Lorentz invariant description of all phenomena with
causes below decks, including electromagnetic phenomena, would be
left. There would still be the fact that the rules would have a Lorentz
invariant behaviour (probably since they are mainly ruled by electromagnetic
forces). The same applies to the clocks, which are all synchronous
to the light clock, i.e. a rule at whose ends an electromagnetic signal
reflects. There would still be the fact that the \emph{reference frames}
may never have a relative speed higher than the light speed in vacuum
(probably as the reference frames are \textquotedblleft{}rigid\textquotedblright{}
rooms, made of rules which, being mainly ruled by electromagnetic
forces, will never be able to move at a speed higher than the one
of the electromagnetic waves). There would still be the fact that
the electromagnetic phenomena do not identify any preferred direction,
that the speed of light is independent of the source motion. I cannot
really see what should change of relativity if we stressed that \emph{PR}
rather than \emph{PRF} should be considered among its foundations.
The possible existence of non Lorentz invariant phenomena is predicted
in \emph{PR} which suggests looking \textquotedblleft{}above decks\textquotedblright{}
in such occurrence. The possible experimental observation of those
non Lorentz invariant effects predicted by the model discussed above
may perhaps challenge other theories explicitly based on \emph{PRF},
which should be revised in this case, and this price should be compared
to the one paid for the renunciation of the local realism. We should,
however, bear in mind that, as already seen, the predictions of the
model aren\textquoteright{}t Lorentz invariant only for markedly exceptional
situations: measurements \textquotedblleft{}of extremely high accuracy\textquotedblright{}
(and geared to the aim) are necessary to be able to identify the non
Lorentz invariant effects predicted by the model. Whether to choose
or not the local realism, accepting its possible consequences, is
just an option. It is not the facts which force us to give up the
local realism. Neither will they be able to force us in the future,
at least as long as we have results consistent with the predictions
of the quantum mechanics, considering that, for adequate $v_{t}$
values, any experiment consistent with the predictions of the quantum
mechanics will be also consistent with the local and realist model
proposed above: 
\begin{quotation}
It should be pointed out that, whether or not such violation of quantum
theory predictions is found experimentally, the model can always be
used by those physicist and engineers who prefer to work with real
localizable entities rather than with the present orthodox concepts
of quantum theory (\citep{Eberhard89} page 172).
\end{quotation}
Naturally, if we assume the above model we will no longer be able
to claim the impossibility to send signals at superluminal speed:
tachyons do send signals at superluminal speed. We could think that
the tachyons are \textquotedblleft{}hidden\textquotedblright{}, inaccessible
to the macroscopic world (I cannot see why we should put forward such
a hypothesis, though); however, in a recent interesting paper, J.-D.
Bancal and al. \citep{Bancal} have shown that \textquotedblleft{}the
models based on hidden influences propagating at a finite speed \emph{v}>\emph{c}\textquotedblright{}
allow \textquotedblleft{}superluminal communication {[}which{]} does
not require access to any hidden physical quantities, but only the
manipulation of measurement devices\textquotedblright{}. Yet, the
possibility to communicate at superluminal speed is considered forbidden
by relativity because of the theorems on causal paradoxes examined
above, which, as we have seen, would fall if the \emph{PRF} weren\textquoteright{}t
assumed. We might claim that to deny the \emph{PRF}, stating the existence
of a privileged reference frame, is against the \textquoteleft{}spirit\textquoteright{}
of relativity, as it is to claim the possibility of superluminal communications.
N. Gisin recently wrote \citep{Gisin}:
\begin{quotation}
Everything looks as if the two parties somehow communicate behind
the scene \citep{Bell}; hence, since they can't communicate at a
speed equal or lower than the speed of light, let's assume they do
so at a speed faster than light. Such an assumption doesn't respect
the spirit of Einstein relativity, {[}...{]}. To define faster than
light hidden communication requires a universal privileged reference
frame in which this faster than light speed is defined. Again, such
a universal privileged frame is not in the spirit of relativity, but
also clearly not in contradiction: for example the reference frame
in which the cosmic microwave background radiation is isotropic defines
such a privileged frame. 
\end{quotation}
Although it seems to me that the approval of the scientific community
on this matter is vast, I take the liberty of disagreeing. In my view
the \textquoteleft{}spirit\textquoteright{} of relativity is elsewhere%
\footnote{In the book published by Schlipp in occasion of Einstein\textquoteright{}s
seventieth birthday, Reichenbach says: \textquotedblleft{}The logical
basis of the theory of relativity is the discovery that many statements,
which were regarded as capable of demonstrable truth or falsity, are
mere definitions'' (\citep{Reichenbach (1949)} p. 293). Among these
\textquotedblleft{}mere definitions\textquotedblright{} a relevant
role is given to synchronization. Einstein answered: \textquotedblleft{}Now
I come to the theme of the relation of the theory of relativity to
philosophy. Here it is Reichenbach's piece of work which, by the precision
of deductions and by the sharpness of his assertions, irresistibly
invites me a brief commentary. {[}...{]} To the question: Do you consider
true what Reichenbach has here asserted, I can answer only with Pilate's
famous question: \textquotedbl{}What is truth\textquotedbl{}?\textquotedblright{}
(\citep{Einstein (1949)} p.676). It is also noteworthy what Einstein
wrote in 1928 in the review \citep{Einstein (1928)} to the book \citep{Reichenbach (1928)}
Reichenbach published in that year: \textquotedblleft{}special care
has been taken to ferret out clearly what in the relativistic definition
of simultaneity is a logically arbitrary decree and what in it is
a hypothesis, i.e., an assumption about the constitution of nature.''
(quoted in \citep{Jammer} pag.181).%
}. That is why I claim that the proposed model does not modify in any
way the relativity: neither its substance, nor its form (the \emph{PR}
is among its postulates, not the \emph{PRF}) nor even its spirit.

\subsection{Causal order}

In my opinion, the model under study poses a question which neither
Eberhard nor Bohm and Hiley have asked. A crucial question would stand
out clearly, at least in my view, were we to accept the model. I will
sum up the issue already presented in other words in \citep{Cocciaro2005}.

The model deprives the light cone of the meaning commonly given. It
is not possible any more to claim that only events inside the light
cone are causally linkable. Besides, once the existence of events
causally linked outside the light cone has been admitted, we have
the immediate consequence that the time order induced by the standard
synchronization is not always consistent with the causal order (this
is exactly the Lorentz non invariance of the model). For instance,
let us assume we have a system of two particles, \emph{A} and \emph{B},
in an entangled state, which is submitted to the measurements $m_{A}$
on \emph{A} and $m_{B}$ on \emph{B}, in the reference frame \emph{R}
of the laboratory. The two measurements are space-like and, given
the standard synchronization of the fixed clocks in the laboratory,
the instants shown on the clocks fixed where (and when) the measurements
are taken are $t_{A}$ and $t_{B}$ respectively. If the two measurements
are correlated according to what predicted by the quantum mechanics,
the model predicts that the two measurements are causally linked:
one measurement is concause to the other. The measurements are causally
ordered: one is \textquotedblleft{}the first\textquotedblright{} and
from the point where the first measurement is taken leaves the tachyon
which will be received by the second particle before the measurement
is taken on it. The point is that the causal order does not necessarily
correspond to the time order resulting from the standard synchronization:
$m_{A}$ might be the first measurement even if we had $t_{A}>t_{B}$.
Such an occurrence does not create any bewilderment once the conventional
character of the standard synchronization (as well as of any other
synchronization) has been clarified \citep{Anderson}; obviously,
however, the model must provide a criterion according to which we
could establish in every circumstance \emph{how to understand} which
of the two measurements is the first. In my opinion, when providing
this criterion, both Eberhard and Bohm and Hiley dodge the crucial
issue expressed above. The model hypothesizes the existence of a preferred
reference frame; through Lorentz\textquoteright{}s transformations
we can identify the instants $t_{A}'$ and $t_{B}'$ shown on the
clocks fixed in the preferred reference frame where (and when) the
two measurements are taken (the clocks of the preferred reference
frames, too, are meant to be synchronized according to a standard
relation). The \textquotedblleft{}first\textquotedblright{} measurement
will be the one associated to the instant $t'$ minor, i.e. if $t_{A}'<t_{B}'$
then the first measurement is $m_{A}$. Essentially both Eberhard
and Bohm and Hiley take the causal order back to the time order resulting
from the standard synchronization in the preferred reference frame.

Finally let us analyse the key issue which is, in my opinion, overlooked.
I agree that the model must give a criterion to establish, \emph{in
principle}, the causal order of the events and that the criterion
given by Eberhard, Bohm and Hiley surely satisfies this condition;
however, let us imagine the experimenter in his laboratory: he takes
the measurements $m_{A}$ and $m_{B}$, sees that the outcomes are
correlated and, accepting the model under study, infers that one has
been the concause of the other. The experimenter does not know the
parameter $\vec{v}$, that is he does not know at what speed or towards
which direction the preferred reference frame is moving with respect
to the Earth. Consequently he cannot know the instants $t_{A}'$ and
$t_{B}'$. Must the experimenter then give up the possibility to know
which of the two measurements was the \textquotedblleft{}first\textquotedblright{}?
Can we believe that something so important as the causal order is
not \textquotedblleft{}written\textquotedblright{} inside some physical
quantity accessible from every reference \emph{R} and nature has decided
to write it only inside some instruments we may or may not decide
to use to describe the events (like the clocks fixed in the preferred
reference frame)? If nobody had put clocks in the preferred reference
frame, would the causal order have been left \textquotedblleft{}written\textquotedblright{}
only in a possibility practicable in principle (and that nobody has
practised)?\\
These are the questions that, in my view, Eberhard, Bohm and Hiley
fail to consider in giving their criterion to establish the causal
order of the events.

I can understand the criticism that could be moved to such questions,
downgrading them to metaphysics and comparing them to wondering how
many angels are able to sit on the point of a needle%
\footnote{From the letter sent by W. Pauli to M. Born on 31 March, 1954:\\
``As O. Stern said recently, one should no more rack one's brain
about the problem of whether something one cannot know anything about
exists all the same, than about the ancient question of how many angels
are able to sit on the point of a needle. But it seems to me that
the Einstein\textquoteright{}s questions are ultimately always of
this kind.'' \citep{Born-Einstein} (pag. 223)%
}, but I believe that in science there is not a definite criterion
which can establish which questions should be considered ridiculous.
Surely what matters in the end is what is \textquotedblleft{}produced\textquotedblright{},
the experimental effects that are expected following our considerations.
Besides, what the experiments say obviously matters. But the reasons
why we make some considerations, the reasons why we consider some
things credible or incredible can be varied. I do not think questions
such as the ones quoted above must necessarily be considered important,
I do believe, however, that they can not be considered ridiculous
precisely as we should not consider ridiculous the question: ``Is
it possible that nature does not work in a realist and local way?\textquotedblright{}
A realist might consider it ridiculous to believe that nature works
in a non local way, nevertheless, in my opinion, he must anyway take
into account theories based on assumptions which appear ridiculous
to him and must then evaluate those theories on the basis of their
predictive ability. Similarly, I think an orthodox should consider
questions ridiculous for him (metaphysical stuff) and evaluate them
on the basis of what they \textquotedblleft{}produce\textquotedblright{}.
The \textquotedblleft{}product\textquotedblright{} I intend to give
in this chapter will be an alternative criterion to understand how
to identify the causal direction (criterion I consider credible since
the one given by Eberhard, Bohm and Hiley seems to me not very credible).

Questions similar to those examined above could be asked also about
subluminal physics. A billiards ball bounces between the cushions.
Without fixing clocks on each cushion and without synchronizing them
according to a definite criterion%
\footnote{Actually it can be easily demonstrated that infinite synchronizations
exist leading to a time order consistent with the causal order.%
}, would we be unable to know which of the two bounces was the concause
of the other, that is, which is the first and which is the second?
Is it possible that something so important as the causal order is
not \emph{written inside the system} under study (i.e: the billiards,
with its cushions and the ball) and it is written only inside measurement
instruments (the clocks fixed on the cushions) which we may even decide
not to use and which also have been synchronized according to a convention
we have chosen ourselves? Without using the clocks, is it impossible
to know which is the cause and which the effect? Naturally some measurements
will have to be taken in order to obtain some physically meaningful
information about the system, some measurement instruments will have
to be used, but the clocks are meters of time intervals, they are
not meters of the causal direction. The criterion followed to establish
the causal direction using clocks is very complicate! We call \emph{CR}
such a criterion, which is the following:

the clock fixed on a cushion measures the time interval $t_{1}$ from
the moment when it was set at the instant 0 to the moment when it
sees the ball bouncing on its cushion, the clock fixed on the other
cushion measures a time interval $t_{2}$ from the moment when it
was set at the instant 0 to the moment when it sees the ball bouncing
on its cushion; the criterion followed to set the clocks at 0 is the
standard synchronization. On the basis of the comparison between the
two time intervals $t_{1}$ and $t_{2}$ it is established which bounce
was the first , that is which of the two caused the other (in the
hypothesis that the two events are causally linked).\\
It is only our unconscious habit to consider time as it had always
stayed in the apriori Olympus which makes the criterion just examined
sound trivial to us. It seems obvious to us that the cause must take
place before the effect and it seems obvious to us that we must give
that \textquotedblleft{}before\textquotedblright{} a meaning linked
to what is shown on the clocks: the cause must be the one taking place
where it is fixed the clock showing an instant less than the one shown
on the clock fixed where the effect will take place. However, getting
time down from the apriori Olympus, relativity teaches us that we
are giving that \textquotedblleft{}before\textquotedblright{} a not
so obvious meaning. A meaning which derives from a complicate criterion
with conventional aspects as well! I also think we should remark the
consideration that it may seem strange that nature has written the
causal direction inside such a complicate criterion.

We have already seen that the criterion \emph{CR} does not work if
the two events are connected causally through the information exchange
occurred at superluminal speed and that this information exchange
would not create causal paradoxes if we decided \emph{not} to assume
\emph{PRF} as one of the foundations of physics. Once \emph{PRF} has
been assumed instead, the impossibility of superluminal signals is
demonstrated (or causal paradoxes would appear) and it is also possible
to demonstrate that the criterion \emph{CR} works regardless of the
reference frame where the events are described and that, with adequate
modifications, it also works independently of the chosen synchronization;
this leads to call, improperly in my opinion, the derived structure
\textquotedblleft{}causal structure\textquotedblright{}. Essentially,
despite the conventionality of the standard synchronization, it is
demonstrated that, assumed the \emph{PRF}, the time order induced
by the standard synchronization is always consistent with the causal
order. At this point the criterion \emph{CR} has the right credentials
and if somebody thinks it is a bit too complicate to be considered
the \textquotedblleft{}true\textquotedblright{} criterion in which
nature has written the causal direction, we can surely reply that
this is his own problem. When I say that the criterion \emph{CR} has
the right credentials, I mean the following: being in the reference
frame \emph{R}, in order to be able to identify the causal order,
it will be necessary to act in a way that might be complicate for
somebody, but it is always only a question of measurements taken inside
\emph{R}. And they are measurements connected with the system under
study. In our example, they are measurements related to the billiards,
the ball; measurements employing clocks fixed in \emph{R}, rules fixed
in \emph{R}, possibly luminous signals (to synchronize the clocks,
should we decide to synchronize them through a luminous signal). The
criterion \emph{CR} does not require anything \textquotedblleft{}odd''.
For instance it does not require us to know where the billiards comes
from, when it was brought to the bar where we are carrying out the
experiment, who the manufacturer is ... It only requires measurements
to be taken on the billiards, using instruments available in \emph{R}.

The situation \emph{changes radically} when we abandon \emph{PRF}
and, as both Eberhard and Bohm and Hiley do, hypothesize the existence
of a preferred reference frame for the propagation of superluminal
signals. We are now still in the reference frame \emph{R} and deal
with our two-particle system in an entangled state. We now examine
the measurements $m_{A}$ and $m_{B}$ on the system under study,
we possibly determine also the time intervals $t_{A}$ and $t_{B}$
shown on the clocks fixed in the place where the measurements are
taken (each of the two intervals starts when the clock is set at 0
and finishes when the corresponding measurement is taken), we determine
the distance between the points where the measurements are taken;
in short, we take all the measurements we might consider useful in
order to know which of $m_{A}$ and $m_{B}$ was the cause and which
the effect, but we realize that the criterion established by both
Eberhard and Bohm and Hiley require us also to take definitely odd
measurements! That criterion says that we first have to determine
$\vec{v}$ (i.e. the speed of the preferred reference frame with respect
to \emph{R}) so that, through Lorentz\textquoteright{}s transformations,
we can then determine the instants $t_{A}'$ and $t_{B}'$ comparing
which we could finally find out which between $m_{A}$ and $m_{B}$
was the cause. It is true that a criterion, to be such, must explain
how to solve the problem in principle (and it is true that in principle
the problem can be solved in the way proposed by Eberhard, Bohm and
Hiley), however, this criterion, in my opinion, \emph{has not} the
right credentials. It would be as if, getting back to the subluminal
world and to our example of the billiards, the criterion \emph{CR},
rather than requiring only measurements taken in \emph{R} on the system
under study, would also require us to know, for example, the manufacturer
of the billiards or the place where it was produced. That is why,
in my opinion, the model presented by Eberhard and Bohm and Hiley
\emph{requires} the research of a \emph{credible criterion}, on the
base of which the causal direction can be established.

The criterion proposed in \citep{Cocciaro2005} can be summarized
as follows, at least in its fundamental aspects:

any time two events are connected causally there is \emph{always}
a \emph{messenger} which propagates from the point where the cause
occurs to the point where the effect takes place. The messenger could
be a particle or, more generally, a signal of any type, subluminal,
luminous or superluminal. A momentum is \emph{always} associated to
the messenger. The direction of the momentum vector of the messenger
is always from the cause to the effect. Given two events causally
connected, in order to identify which is the cause and which the effect,
we measure the momentum of the messenger.

\subsection{Criterion of causality and time ordering.}

In a given reference frame \emph{R} we consider two events associated
to the quadrivectors $\left(ct_{1},\vec{x}_{1}\right)$ and $\left(ct_{2},\vec{x}_{2}\right)$
respectively. It will be possible to associate to the two events the
quadrivector $\Delta x_{\mu}\equiv\left(\Delta ct=ct_{2}-ct_{1},\Delta\vec{x}=\vec{x}_{2}-\vec{x}_{1}\right)$,
which, like all quadrivectors, enables us to define the corresponding
invariant:
\[
\left(\Delta x_{\mu}\cdot\Delta x_{\mu}\right)\equiv\left(\Delta ct\right)^{2}-\left(\Delta\vec{x}\right)^{2}.
\]
It can be easily shown that $\left(\Delta x_{\mu}\cdot\Delta x_{\mu}\right)>0\Leftrightarrow\nicefrac{\left(\Delta\vec{x}\right)^{2}}{\left(\Delta ct\right)^{2}}<1$,
that is the sign of the invariant $\left(\Delta x_{\mu}\cdot\Delta x_{\mu}\right)$
can discriminate the pairs of events connectible through subluminal
signals from those not connectible through subluminal signals: in
the former case the invariant is positive, in the latter it is negative.
Let\textquoteright{}s now limit the analysis to the case $\left(\Delta x_{\mu}\cdot\Delta x_{\mu}\right)>0$
and hypothesize that the two events are linked by a causal relation
which could be easily identified. For instance the event $\left(ct_{1},\vec{x}_{1}\right)$
corresponds to the creation of a certain particle (cause) and the
event $\left(ct_{2},\vec{x}_{2}\right)$ corresponds to the detection
of the particle (effect). In accordance with how it has been defined,
the vector $\Delta\vec{x}$ heads from the cause to the effect. However,
we could hypothesize a case in which it is not known if the particle
has been generated in $\vec{x}_{1}$ or in $\vec{x}_{2}$ (i.e. if
$\Delta\vec{x}$ must be defined as $\vec{x}_{2}-\vec{x}_{1}$ or
$\vec{x}_{1}-\vec{x}_{2}$), in such a case we could turn to the reading
of the instants $t_{1}$ and $t_{2}$ to identify the causal direction.
It can be easily shown that, if the events can be connected through
particles in subluminal motion, the cause must be associated to the
event with the lower instant. Indeed, assuming that the particle is
generated in $\vec{x}_{1}$ and detected in $\vec{x}_{2}$, under
the standard synchronization, we will have that the clock in $\vec{x}_{2}$
will indicate the instant $\bar{t}=t_{1}+\nicefrac{\left|\Delta\vec{x}\right|}{c}$
exactly when it is reached by a light beam leaving $\vec{x}_{1}$
simultaneously with the creation of the particle. Moving subluminally,
the particle will reach the point $\vec{x}_{2}$ after the light beam
, that is when the clock set in that point will indicate $t_{2}>\bar{t}$.
Having synchronized the clocks fixing $\bar{t}>t_{1}$, it will follow
$t_{2}>t_{1}$.\\
We could, however, identify the causal direction without turning to
the reading of the instants $t_{1}$ and $t_{2}$. An energy-momentum
quadrivector is associated to the particle and we could define it
as
\begin{equation}
p_{\mu}=\left(E,\vec{p}\right)\equiv\sqrt{\frac{\left(p_{\mu}\cdot p_{\mu}\right)}{\left(\Delta x_{\mu}\cdot\Delta x_{\mu}\right)}}\Delta x_{\mu}.\label{eq:7}
\end{equation}
The \eqref{eq:7} is to be understood as follows: $p_{\mu}$ is proportional
to $\Delta x_{\mu}$, according to a positive constant which will
necessarily correspond to the root of the ratio between the invariants
$\left(p_{\mu}\cdot p_{\mu}\right)$ and $\left(\Delta x_{\mu}\cdot\Delta x_{\mu}\right)$.
Particularly from \eqref{eq:7} it follows that the vector momentum
associated to the particle will be defined as
\[
\vec{p}\equiv\sqrt{\frac{\left(p_{\mu}\cdot p_{\mu}\right)}{\left(\Delta x_{\mu}\cdot\Delta x_{\mu}\right)}}\Delta\vec{x},
\]
that is $\vec{p}$ is by definition concordant to the causal direction
(i.e. its direction is the same as the direction of $\Delta\vec{x}$).
This means that in case we did not know if the cause was in $\vec{x}_{1}$
or in $\vec{x}_{2}$ (that is if the correct definition of $\Delta\vec{x}$
was $\vec{x}_{2}-\vec{x}_{1}$ or $\vec{x}_{1}-\vec{x}_{2}$) we could
resort to measuring the vector $\vec{p}$ (particularly to measuring
the direction of that vector) to identify the causal direction.\\
We can observe, as noticed above, that the fact that the sign of $\Delta ct=ct_{2}-ct_{1}$
indicates the causal direction results from the definition of $\Delta ct$
(e.g. from the assumed synchronization) as well as from the fact that
the chosen events are connectible through subluminal motion, namely
that $\left(\Delta x_{\mu}\cdot\Delta x_{\mu}\right)>0$. Therefore
we should not be surprised if for $\left(\Delta x_{\mu}\cdot\Delta x_{\mu}\right)<0$,
that is, for pairs of events connectible through superluminal motion,
the sign of $\Delta ct$ induced by the standard synchronization could
fail to indicate the causal direction. Indeed, assuming that in $\vec{x}_{1}$
a tachyon is generated, which will later be detected in $\vec{x}_{2}$,
even if we had the \textquotedblleft{}right\textquotedblright{} sign
for $\Delta ct=ct_{2}-ct_{1}$ (i.e. was $\Delta ct>0$) in \emph{R},
we could always find appropriate reference frames $R'$, moving with
velocity $\vec{\beta}c$ with respect to \emph{R}, for which will
be $\Delta ct'=ct'_{2}-ct'_{1}<0$. Indeed Lorentz transformations
give
\[
\Delta ct'=\frac{1}{\sqrt{1-\vec{\beta}^{2}}}\left(\Delta ct-\vec{\beta}\cdot\Delta\vec{x}\right)
\]
and being $\left(\Delta x_{\mu}\cdot\Delta x_{\mu}\right)<0\Leftrightarrow\left|\nicefrac{\Delta\vec{x}}{\Delta ct}\right|>1$
it will be always possible to determine appropriate $\vec{\beta}$,
with $\left|\vec{\beta}\right|<1$, so to make $\left(\Delta ct-\vec{\beta}\cdot\Delta\vec{x}\right)<0$.
As often stated above, the time ordering induced by the standard synchronization
can not be always (i.e. for every reference frame) concordant with
the causal ordering. In my opinion, we should always accept without
doubt the fact that the causal ordering of events must be totally
independent of the reference frame chosen to describe the same events.
Nevertheless vast literature exists on the so-called \textquotedblleft{}principle
of reinterpretation\textquotedblright{} \citep{Bilaniuk} according
to which cause and effect could reverse their roles depending on the
reference frame in which the events are described. In our example,
according to the reference frame \emph{R} the tachyon is emitted in
$\vec{x}_{1}$ and detected in $\vec{x}_{2}$, according to the reference
frame $R'$ the same tachyon would be emitted in $\vec{x}'_{2}$ and
detected in $\vec{x}'_{1}$. This in order to \textquotedblleft{}preserve\textquotedblright{}
the time ordering, to admit the possibility of superluminal signals
while continuing to claim that all types of signals are generated
\textquotedblleft{}temporally before'' being detected; meaning with
temporally before the time ordering induced by the standard synchronization.
Here we strongly object to this presentation and simply accept the
fact that the time ordering is not a criterion which allows to identify
the causal direction in the case of superluminal signals.\\
We have seen above that for subluminal motions we could identify the
causal direction even without resorting to the time ordering: we could
simply measure the momentum $\vec{p}$ of the particle propagating
from the point where the cause occurs to the point where the effect
takes place. According to the local realism we can hypothesize that
there is \emph{always}, for every pair of events linked by a causal
relation, a \textquotedblleft{}messenger\textquotedblright{} propagating
from the point where the cause occurs to the point where the effect
occurs. We hypothesize that for superluminal signals it is possible
to identify an energy-momentum quadrivector in the same way as for
the subluminal particles, namely according to%
\footnote{We can notice that from \eqref{eq:7}, it appears that $\left(\Delta x_{\mu}\cdot\Delta x_{\mu}\right)$
and $\left(p_{\mu}\cdot p_{\mu}\right)$ have the same sign so the
radicand in \eqref{eq:7} will be always defined. Consequently for
the superluminal signals also the invariant $\left(p_{\mu}\cdot p_{\mu}\right)$
will be negative.%
} \eqref{eq:7}. The criterion proposed to determine the causal direction
of events causally connected is to measure the direction of the momentum
vector of the messenger propagating from the point where the cause
occurs to the point where the effect takes place.

\section{Conclusions}

The realist and local model proposed by Eberhard in 1989 and presented
again by Bohm and Hiley in 1993 is perfectly consistent with relativity
once stressed that this must be based on the principle of relativity
as originally expressed by Galileo. All known experimental evidence
are consistent with the model.

As it is known, Einstein\textquoteright{}s corpse was cremated and
the ashes dispersed in an unknown place. Thus his grave does not exist;
however, quoting the figure of speech used above by Davies, Brown
and Bell, we could say that observing the scientific community pay
scarce attention to a realist and local model, consistent with all
known experimental evidence and with relativity, this would definitely
make somebody turn in his grave: Einstein.

\section*{Acknowledgements}

I heartily thank E. Fabri for his help and suggestions, particularly
in pointing out Galileo's key passage which constitute the core of
the whole work. I have merely drawn what I consider the necessary
consequences from his teachings.\\
I would also like to thank P. H. Eberhard, S. Faetti, L. Fronzoni,
N. Gisin and V. Moretti, for their valuable contributions and criticisms
during the elaboration of the work.\\
Finally my thanks to CM for her care in translating my paper.

\end{document}